\documentclass[aps,pra,amsmath,amssymb,twocolumn,preprintnumbers,superscriptaddress,10pt,longbibliography]{revtex4-2}

\usepackage[utf8]{inputenc}                                             

\usepackage{xcolor}                                                     

\usepackage{amsmath,amssymb}                                            
\usepackage{amsthm}                                                     
\usepackage[italicdiff]{physics}                                        
\usepackage{bm}                                                         
\usepackage{tensor}                                                     
\usepackage{siunitx}                                                    
\usepackage{simplewick}                                                 

\usepackage{graphicx}                                                   
\usepackage{tikz}														
\usepackage{pgfplots}                                                   
\pgfplotsset{compat=1.14}
\usetikzlibrary{spy}

\usepackage{hyperref}                                                   

\renewcommand*{\vec}[1]{\bm{\mathrm{#1}}}                               
\newcommand*{\reals}{\mathbb{R}}                                        
\newcommand*{\lp}{\mathopen{}\left}                                     
\newcommand*{\rp}{\right}                                               
\newcommand*{\hc}{\text{h.c.}}                                          

\renewcommand*{\ket}[1]{\left| #1 \right\rangle}                        
\renewcommand*{\bra}[1]{\left\langle #1 \right|}                        
\renewcommand*{\braket}[2]{\left\langle #1 \middle| #2 \right\rangle}   
\newcommand*{\exv}[1]{\left\langle #1 \right\rangle}                    
\newcommand*{\norder}[1]{
    :\hspace{1pt}\mathrel{#1}\hspace{1pt}:\hspace{2pt}%
}

\newcommand*{\tbuff}{\mathchoice{\quad}{\>}{\>}{\>}}                    
\newcommand*{\kbuff}{\:}                                                

\newcommand*{\fref}[1]{Fig.~\ref{#1}}                                   

\begin{document}


\title{Strictly localized three-dimensional states close to single photons}

\author{Kai Ryen}
\affiliation{Department of Physics, University of Oslo, NO-0316 Oslo, Norway}

\author{Jan Gulla}
\affiliation{Department of Technology Systems, University of Oslo, NO-0316 Oslo, Norway}

\author{Johannes Skaar}
\email{johannes.skaar@fys.uio.no}
\affiliation{Department of Physics, University of Oslo, NO-0316 Oslo, Norway}

\date{\today}

\begin{abstract}
A class of strictly localized states which can be made arbitrarily close to single photons is constructed, and expressions for central properties are provided. It is demonstrated that single photon states can be well approximated by these states down to localization scales on the order of a few pulse cycles. The results readily generalize to states close to $n$-photons. We also provide upper and lower bounds for the fidelity between a given single photon and any state strictly localized to a given volume. These results constitute the limit of photon localization, complementary to the weak-localization limit (I. Bialynicki-Birula, Phys. Rev. Lett. 80, 5247 (1998)).

\end{abstract}


\maketitle

\section{Introduction}
The frequency restrictions on single photon states prohibit them from being perfectly localized to a finite region \cite{knight1961, hegerfeldt1998causality}, which raises the question of the degree to which they can be localized. Efforts into finding progressively finer bounds on this ``weak localization" have a long tradition \cite{newton1949, mandel1966, jauch1967, amrein1969, mandel1995, adlard1997, bialynicki-birula1998, saari2005photon}. A statement in the famous textbook by Mandel and Wolf \cite{mandel1995} claims that the energy density of a single photon is spread out over space asymptotically like $r^{-7}$, where $\vec r$ is the position, a result obtained in the 1960s \cite{jauch1967, amrein1969}. Later, better localized photons were found \cite{adlard1997}. In 1998, Bialynicki-Birula argued that the limit of localization is near-exponential \cite{bialynicki-birula1998}, as dictated by the Paley-Wiener criterion \cite{paley1934}.

It is natural to consider the limits of localization from another point of view. For states which are \emph{strictly localized} to a certain volume, how close can they be to a single photon? Strictly localized states are important because they correspond to those generated by on-demand sources: If an experimentalist pushes a button, the resulting state must be indistinguishable from vacuum outside the button's light cone in order to preserve causality \cite{gulla2020,gulla2021b}.

As defined by Knight \cite{knight1961}, a state $\ket{\psi}$ is strictly localized to a spacetime volume if
\begin{equation}
    \bra{\psi} L \ket{\psi} = \bra{0} L \ket{0}
\end{equation}
for all local operators $L$ (which consist of sums and products of the field operator, including derivatives and integrals), evaluated \emph{outside} of the volume. Here $\ket{0}$ is the vacuum state. Licht published a more comprehensive addition to the theory of strict localization \cite{licht1963}, showing that for each strictly localized state $\ket{\psi}$ there exists a unique bounded operator $W$, which is partially isometric \footnote{Despite \eqref{eq:partiso}, a Licht operator $W$ is not necessarily unitary.},
\begin{subequations}\label{eq:lichtdef}
\begin{equation}\label{eq:partiso}
    W^\dagger W = 1,
\end{equation}
commutes with all local operators $L$ outside the state's localization volume,
\begin{equation}\label{eq:lichtop}
    [W, L] = 0,\ \ \ \ 
\end{equation}
and excites the strictly localized state from vacuum,
\begin{equation}
    W\ket{0} = \ket{\psi}.
\end{equation}
\end{subequations}
The first two conditions are sufficient to guarantee that the latter state is strictly localized. We will refer to $W$ as a \emph{Licht operator}.

We have previously constructed Licht operators and associated states arbitrarily close to single photons, with strict one-sided localization in time \cite{gulla2020}. These states are useful for establishing the limits of on-demand one-dimensional photon sources \cite{gulla2020, gulla2021b}, and for establishing strict causality in tunneling experiments \cite{gulla2021}.

The states in \cite{gulla2020, gulla2021b} cannot, however, be used to determine the limit of localization in three-dimensional space. The key to understanding the localization limit is that associated with the annihilation operators there can be any wavevector but only positive frequencies (see the standard expression \eqref{eq:Efieldop} below). This asymmetry between space and time makes it nontrivial to go from localization in time to localization in space, except for forward-propagating modes in 1D. 

In this work we first construct a class of three-dimensional states close to single photons, strictly localized to finite, causally expanding volumes. This is done in three steps: In Sec. \ref{sec:class_loc} we construct a localized, classical scalar wave in a suitable form, as a superposition of two modes. In Sec. \ref{sec:qstates}, with the help of a nontrivial assumption \eqref{eq:CEsupport} we obtain photonic, quantum states with the desired localization property. In Sec. \ref{sec:constructvecf} we prove that assumption \eqref{eq:CEsupport} can be fulfilled, as a single-curl construction related to the localized, classical scalar wave.

With the class of states in hand, we determine closed-form expressions for the most important properties, like the expectation value of normal-ordered squared fields, and fidelity with a single photon, and give numerical examples (Sec. \ref{sec:etastates} and Appendix \ref{sec:etaproperties}). Applying powers of the corresponding Licht operator leads to states close to $n$-photon number states. Finally, we find upper and lower bounds for the fidelity between a given single photon and any state strictly localized to a given volume, expressed from the photon's spatial tail outside the localization volume (Sec. \ref{sec:upper_bound}). Together with the fidelity expression this constitutes the limit of photon localization, complementary to Bialynicki-Birula's weak localization limit \cite{bialynicki-birula1998}.

\section{Discrete modes}
The $\vec E$-field operator is \cite{cohen-tannoudji1997}
\begin{equation}\label{eq:Efieldop}
    \vec E(\vec r, t) = \sum_{i=1}^2\int \mathcal{E}(\omega) \vec e_i(\vec k)  a_i(\vec k) e^{i\vec k \cdot \vec r - i\omega t}d^3k +\hc,
\end{equation}
where $\vec e_i$ is a polarization vector such that $\vec e_1^* \cdot \vec e_2 = \vec k \cdot \vec e_i = 0$ (typically either linearly or circularly polarized), and $\mathcal{E}(\omega) \propto \sqrt{\omega}$, where $\omega=kc$, is the usual normalization factor. The operator may be expanded in a discrete (countable) basis \cite{tatarskii1990} in the following way: suppose we have two orthonormal complete bases $\{\xi_{ni}\}_n$ for $i \in \{1, 2\}$ such that
\begin{align*}
    &\int \ \xi^*_{ni}(\vec k) \xi_{mi}(\vec k) d^3k= \delta_{n m} \\
    &\sum_n \xi^*_{ni}(\vec k) \xi_{ni}(\vec k^\prime) = \delta(\vec k - \vec k^\prime),
\end{align*}
for each polarization index $i$ respectively. The bases may be equal or different. In this decomposition the field operator is written as
\begin{align} \label{eq:pulsemodeEop}
    \vec E(\vec r, t) = \sum_n \sum_{i=1}^2 \vec E_{ni}(\vec r, t) a_{ni} + \hc,
\end{align}
where we have defined the electric field modes
\begin{equation}
    \vec E_{ni}(\vec r, t) = \int \ \vec e_i(\vec k) \mathcal{E}(\omega) \xi_{ni}(\vec k) e^{i\vec k \cdot \vec r - i\omega t}d^3k, \label{eq:Emode}
\end{equation}
and the associated ladder operators
\begin{equation}
    a_{ni}^\dagger = \int \ \xi_{ni}(\vec k) a_i^\dagger(\vec k) d^3k,
\end{equation}
satisfying
\begin{subequations}
\begin{align}
    [a_{ni}, a_{mj}] &= [a^\dagger_{ni}, a^\dagger_{mj}] = 0, \\
    [a_{ni}, a^\dagger_{mj}] &= \delta_{nm}\delta_{ij}.
\end{align}
\end{subequations}
The states produced by the creation operators follow the standard formalism, e.g.,
\begin{align}
    a_{12}^\dagger\ket{k_{11}, l_{12}, \ldots} = \sqrt{l+1}\ket{k_{11}, (l+1)_{12}, \ldots},
\end{align}
for integer $k$, $l$, $\ldots$ Here the subscripts $11$ and $12$ etc. correspond with the modes excited by the identically indexed ladder operators.

\section{Strictly localized states close to single photons}\label{sec:etastates}

\subsection{Localized classical scalar waves}\label{sec:class_loc}
Perfect single photons cannot be strictly localized due to their positive-frequency spectrum (or energy) \cite{knight1961,hegerfeldt1998causality,bialynicki-birula1998,gulla2020}. On the other hand, classical wave functions are straightforward to localize to any extent, since there are no restrictions on their spectra.

To prepare for the construction of non-classical, strictly localized states, we will first describe classical, localized wave functions with causally evolving support. These will turn out to be useful for obtaining our quantum states. A source has been active for $t<0$, and we consider $t>0$ after which the source has been turned off.

One may think of a spherical symmetric wave propagating outwards as a 3D analogue of a forward propagating mode in 1D. However, this is in fact not true; an outward propagating wave is not a true mode (see \eqref{eq:phi}-\eqref{eq:phi_simpl} below). Consequently the spacetime dependence will not be of the form $r-ct$ as would be the 3D analogue of $x-ct$ in 1D. In addition, there is another complication related to Birkhoff's theorem, which says that are no spherical symmetric electromagnetic waves.

\begin{figure}[tb]
    \begin{tikzpicture}
    \draw[fill, fill opacity=0.16] (2, 2) circle (0.4cm);
    \draw (2, 3.0) node {$V(0)$};
    \draw[fill, fill opacity=0.16] (6, 2) circle (1.2cm);
    \draw (6, 3.8) node {$V(t)$};

    \draw[color=red] (2, 2) -- (2.4, 2);
    \draw[color=red] (2.1, 2.15) node {$r_0$};

    \draw[color=red] (6, 2) -- (6.4, 2);
    \draw[color=red] (6.1, 2.15) node {$r_0$};
    \draw (6.40, 2.15) node {$+$};
    \draw[color=blue] (6.4, 2) -- (7.2, 2);
    \draw[color=blue] (6.7, 2.15) node {$ct$};

    \draw[dashed] (2, 2.4) -- (6, 3.2);
    \draw[dashed] (2, 1.6) -- (6, 0.8);
    
    \draw[->] (1.2,0.5) -- (7.6,0.5);
    \draw (7.2, 0.2) node {$t$};
    \end{tikzpicture}
    \caption{\label{fig:CEV}
    A causally expanding sphere is characterized by a radius evolving as $r_0+ct$. A general causally expanding volume can be written as a union of such spheres with different initial radii and centers, in more rigorous terms a union of space-like slices of forward light cones.
    }
\end{figure}
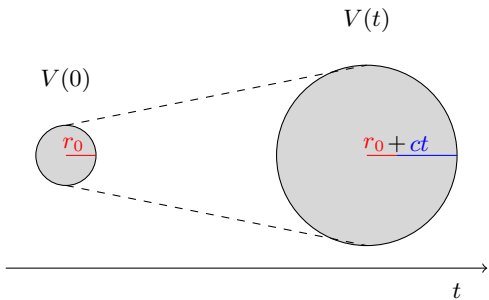

Consider a spherically symmetric scalar wave function in the form
\begin{equation}
    \phi(r, t) = \int \zeta(k) e^{i \vec k \cdot \vec r - i\omega t} d^3k, \tbuff t \geq 0, \label{eq:phi}
\end{equation}
with the dispersion relation $\omega = ck$. By performing the angular integrals, we obtain
\begin{equation}
    \phi(r, t) = -\frac{2\pi i}{r} \lp[ u(r-ct) - u(-r-ct) \rp], \label{eq:phi_simpl}
\end{equation}
where
\begin{equation}\label{eq:uint}
    u(r) = \int_0^\infty \zeta(k)k e^{ikr}\dd k.
\end{equation}
The Paley-Wiener criterion \cite{paley1934} means that $u(r)$ cannot be finitely supported since only positive wavenumbers $k$ are involved in the Fourier integral \eqref{eq:uint}. Therefore $\phi(r,t)$ cannot be finitely supported except at isolated times. However we may get around this by defining the function
\begin{equation}
    f(r, t) = \phi_1(r, t)+\phi^*_2(r, t), \label{eq:f}
\end{equation}
where $\phi_1(r,t)$ and $\phi_2(r,t)$ are of the type in \eqref{eq:phi}, with associated $\zeta_1(k)$ and $\zeta_2(k)$, respectively.  We can write \eqref{eq:f} as a combination of d'Alembert solutions \footnote{These solutions contain values for negative $r$; this regime should of course be neglected. Evolving the solutions in time one may observe two important aspects: Firstly the inward and outward propagating terms cancel at the origin, ensuring that the solution is finite despite $r$ in the denominator. Secondly the inwards term completes the solution for a free field; this part converges to the origin and disappears as the outwards term appears.}
\begin{equation}\label{eq:fg}
    f(r, t) = -\frac{2\pi i}{r} \lp[ g(r-ct)-g(-r-ct) \rp].
\end{equation}
Here we have defined the \emph{seed function}
\begin{equation}
    g(r) = \int_{-\infty}^\infty G(k) e^{ikr} dk, \label{eq:g}
\end{equation}
where
\begin{equation} \label{eq:Gdef}
    G(k) = \begin{cases}
        \zeta_1(k) k; &k\geq 0, \\
        \zeta_2^*(-k) k; &k<0.
    \end{cases} 
\end{equation}
With a suitable choice of $\zeta_1(k)$ and $\zeta_2(k)$, we may obtain $g(r)$ with finite support $[0, r_0)$, because the Fourier integral contains (almost) all wavenumbers. Then
\begin{equation}
    f(r, t) = 0; \ \ \ \ \vec r \notin V(t) \label{eq:CEsupport_radial_scalar}
\end{equation}
for the \emph{causally expanding volume} $V(t) = \{\vec r: \ \ |\vec r| < r_0 + ct, \ \ t \geq 0\}$ (see Fig. \ref{fig:CEV}). 

In this way, by defining a finitely supported seed function $g(r)$ we will fully specify a spherically symmetric, localized and causally evolving wave equation solution $f(r, t)$. Note that the positive and negative signs of $k$ in \eqref{eq:Gdef} do not correspond to outward and inward propagation, but rather to the two different modes $\phi_1(r,t)$ and $\phi_2(r,t)$, which both contains inward and outward components.

We can now in principle compose various more complicated functions by superposition. Given a linear combination
\begin{equation}
    f_\Sigma(\vec r, t) = \sum_{i=0}^N f_i(|\vec r-\vec r_i|, t)
\end{equation}
of functions satisfying \eqref{eq:CEsupport_radial_scalar} for various causally expanding spheres (which may have different initial radii and centers), $f_\Sigma(\vec r,t)$ will be a solution supported on their union. For simplicity, in this work we will restrict ourselves to the spherically symmetric solutions \eqref{eq:fg}.

\subsection{Defining the states}\label{sec:qstates}
Analogously to \eqref{eq:f}, in order to construct a state strictly localized to a causally expanding volume $V(t)$ for $t\geq 0$, we require two modes. We will use the electric field modes $E_{1i}(\vec r, t)$ and $E_{2i}(\vec r, t)$, with polarizations $i$ and $j$, respectively. From these we define the quantity
\begin{equation}
    \vec f(\vec r, t) = \vec E_{1i}(\vec r, t) + \frac{\vec E_{2j}^*(\vec r, t)}{C}, \label{eq:fdef}
\end{equation}
where $C>1$ is a constant. Supposing this function satisfies
\begin{equation}
    \vec f(\vec r, t) = 0; \ \ \ \ \vec r \notin V(t), \label{eq:CEsupport}
\end{equation}
then the electric field operator \eqref{eq:pulsemodeEop} will take the form
\begin{equation} \label{eq:Eoutside}
    \vec E(\vec r, t) = \vec E_{1i}(\vec r, t) (a_{1i} - C a_{2j}^\dagger) + \hc + \ldots
\end{equation}
when evaluated outside the volume $V(t)$. Here terms with (mode index, polarization) different from $(1,i)$ and $(2,j)$ are implied by the lower dots. Since we may freely choose the combination of polarization indices $i,\ j$ without affecting the rest of this derivation, they are omitted in the following.

We introduce the partially isometric operator
\begin{equation} \label{eq:Wdagger}
    W \equiv S^\dagger A_1^\dagger S,
\end{equation}
where $A_1^\dagger = a_1^\dagger\frac{1}{\sqrt{ a_1 a_1^\dagger}}$ is the step operator for the first mode, and $S = e^{\gamma (a_1 a_2-a_1^\dagger a_2^\dagger)}$ is a unitary two-mode squeeze operator \cite{schumaker1985new,gulla2020}, with the squeezing parameter $\gamma$ satisfying $\tanh \gamma = 1/C$. The squeeze operator satisfies
\begin{equation}        
    S (a_1-Ca_2^\dagger) S^\dagger = -a_2^\dagger \sqrt{C^2-1}, 
\end{equation}
which means that
\begin{equation}        
    [W, \ a_{1} - C a_{2}^\dagger] = 0.
\end{equation}
Likewise $S(a_1^\dagger-Ca_2)S^\dagger = -a_2 \sqrt{C^2-1}$, and so $W$ commutes with the Hermitian conjugate in \eqref{eq:Eoutside} as well. Thus condition \eqref{eq:CEsupport} implies the key property
\begin{equation}
    [W, \vec{E}(\vec r, t)] =0; \ \ \ \ \vec r \notin V(t).
\end{equation}
This result readily extends to operator products, sums and derivatives of $\vec E$ evaluated outside of $V(t)$, so we conclude that \eqref{eq:Wdagger} is by definition \eqref{eq:lichtop} a Licht operator producing a strictly localized state
\begin{equation}
    \ket{\eta_{1,2}} \equiv W \ket{0}. \label{eq:etastate}
\end{equation}
We characterize this state by the parameter
\begin{equation}
    \eta = \frac{1}{C^2+1}. \label{eq:eta}
\end{equation}
From the definition of $W$ the state must be of the form
\begin{equation}\label{eq:r_state_expansion}
    \ket{\eta_{1,2}} = c_1 \ket{1_1 \kbuff 0_2} + c_2 \ket{2_1 \kbuff 1_2} + c_3 \ket{3_1 \kbuff 2_2} + \dotsb,
\end{equation}
for coefficients $c_1, c_2, \dotsc$ The state's closeness to a single photon is given by the fidelity (see Appendix \ref{sec:etaproperties})
\begin{align}
    F &\equiv \abs{\braket{1_1\ 0_2}{\eta_{1,2}}} = \sqrt{\frac{(1-2\eta)^{3}}{\eta^2-\eta^3}}\text{Li}_{-1/2}\lp(\frac{\eta}{1-\eta}\rp), \label{eq:Fid} \\
    &= 1-(3/2-\sqrt{2})\eta + O(\eta^2) \approx 1-0.09\eta, \label{eq:Fid1ord}
\end{align}
where $\text{Li}_{-1/2}(x)$ is the polylogarithm function of order $-1/2$. For small $\eta$ we can use the first order expansion \eqref{eq:Fid1ord}.

The parameter $\eta$ quantifies the degree to which $\xi_2(\vec k)$ contributes to the state \eqref{eq:etastate}. In Sec. \ref{sec:numexample} it will become clear that by having a sufficiently large support region for $\vec f(\vec r,t)$, we can obtain $\eta$ as low as we want. Then $F \to 1$ and the state approaches a single photon state. This is intuitive; the closer $\vec f(\vec r, t)$ is to the single photon mode $\vec E_1(\vec r, t)$, the closer $\ket{\eta_{1,2}}$ is to a single photon state.

The strategy of our construction is then to find a function $\vec f(\vec r, t)$ obeying \eqref{eq:CEsupport}, and which can be decomposed into the form \eqref{eq:fdef}. Recalling definition \eqref{eq:Emode}, the latter condition amounts to determining $C$ and normalizing $\vec f(\vec r, t)$ such that 
\begin{equation}
    \int |\xi_1(\vec k)|^2 \dd^3 k = \int |\xi_2(\vec k)|^2 \dd^3 k = 1.
\end{equation}
We must also ensure that
\begin{equation}
    \int \xi^*_2(\vec k)\xi_1(\vec k) d^3k = 0
\end{equation}
is satisfied if both modes belong to the same polarization, since they are members of the same orthogonal basis.
The constant $C$ and basis functions $\xi_1(\vec k)$ and $\xi_2(\vec k)$ then fully characterize a solution. We say that $\vec f(\vec r, t)$ \emph{generates} the Licht operator \eqref{eq:Wdagger} and its strictly localized state \eqref{eq:etastate}.

The natural extension of the above results is to consider powers of the Licht operator. Indeed, compositions of Licht operators are also Licht operators for the union of their localization volumes. This follows directly from \eqref{eq:lichtdef}. For small $\eta$, $W^n$ therefore produces a strictly localized state close to an $n$-photon state, which we label $\ket{\eta_{1,2}^n}$. Squared field and number expectation values for $\ket{\eta_{1,2}^n}$, as well as the  fidelity $|\braket{n_1, 0_2}{\eta_{1,2}^n}|$, are derived in Appendix \ref{sec:etaproperties}.

\subsection{Constructing the generating function}\label{sec:constructvecf}
The generating function $\vec f(\vec r,t)$ solves the wave equation component-wise. We can therefore construct it from a scalar function similarly to how the electric or magnetic fields are related to a one-component vector potential. The curl operators will preserve the original function's support in space and time, and ensure transversality of the result. We choose
\begin{equation}
    \vec f(\vec r, t) = -i \nabla \times \lp[ f(r, t) \vec{\hat z} \rp], \label{eq:singlecurl}
\end{equation}
where $f(r,t)$ is given by \eqref{eq:f}, and $\vec{\hat z}$ is the unit vector in the $z$-direction. We can find an expression for $\vec f(\vec r,t)$ by substituting \eqref{eq:fg}-\eqref{eq:Gdef} into \eqref{eq:singlecurl}. However, we want to express $\vec f(\vec r,t)$ in the form \eqref{eq:fdef} to identify basis functions $\xi_1(\vec k)$ and $\xi_2(\vec k)$. To this end, we use \eqref{eq:f} with \eqref{eq:phi}: 
\begin{equation}\label{eq:fGdecomp}
    \begin{aligned}
        \vec f(\vec r, t) = \int &\vec e_1(\vec k) \sin \theta G(k) e^{i\vec k \cdot \vec r - i\omega t} d^3k \\
        &+ \int \vec e_1(\vec k) \sin \theta G(-k) e^{-i\vec k \cdot \vec r + i\omega t} d^3k,
    \end{aligned}
\end{equation}
where $\theta$ is the angle between $\vec{\hat z}$ and $\vec{\hat k}=\vec k/k$, and $\vec e_1(\vec k) = \frac{\vec{\hat k} \times \vec{\hat z}}{\sin \theta}$ is a linear polarization vector. Indeed, \eqref{eq:fGdecomp} is in the form \eqref{eq:fdef} for two linearly polarized modes,
\begin{equation}
    \vec f(\vec r, t) = \vec E_{11}(\vec r, t) + \frac{\vec E_{21}^*(\vec r, t)}{C},
\end{equation}
corresponding to basis functions $\xi_1(\vec k)$ and $\xi_2(\vec k)$:
\begin{subequations}\label{eq:singlecurlxi}
    \begin{align}
        &\xi_1(\vec k) = \sin \theta \frac{G(k)}{\mathcal{E}(\omega)}, \label{eq:singlecurlxi1}\\
        &\frac{\xi_2(\vec k)}{C} = \sin \theta \frac{G^*(-k)}{\mathcal{E}(\omega)}.
    \end{align}
\end{subequations}
By scaling $G$ both $\xi_1(\vec k)$ and $\xi_2(\vec k)$ can be normalized, provided $G(k)$ falls off sufficiently rapidly. This determines $C$, so \eqref{eq:eta} becomes
\begin{equation}\label{eq:etaG}
\eta = \frac{\int_{-\infty}^{0} |G(k)|^2|k|\dd k}{\int_{-\infty}^{\infty} |G(k)|^2|k|\dd k}.
\end{equation}
If $G(k)$ is picked with main weight for $k>0$, we obtain $\eta < 1/2$, which by \eqref{eq:Fid1ord} means that our state $\ket{\eta_{1,2}}$ gets close to a single photon.

So far there is no guarantee that the above basis functions will be orthogonal (which is required in this case, since they are associated with identical polarization). In the absence of tremendous luck, however, one can always construct such a pair by modifying a given seed function \eqref{eq:g} according to the orthogonalization procedure detailed in Appendix \ref{sec:orton}. It turns out that the orthogonalization always leads to improved fidelity compared to the unmodified case. In other words, in the neighborhood of an arbitrary seed function, there will always exist another seed function which leads to orthogonal modes, without any loss of fidelity. The orthogonalization procedure takes us to that function in one straightforward step.

We can now summarize the algorithm for obtaining a strictly localized state close to a single photon:
\begin{enumerate}
    \item Pick a seed function $g(r)$ with support in the desired region $0 \leq r < r_0$, whose Fourier transform $G(k)$ has its main weight for $k>0$. Make sure that $g(r)$ is sufficiently smooth, such that $\xi_1$ and $\xi_2$ are normalizable in 3D.
    \item Modify $G(k)$ using the orthogonalization procedure in Appendix \ref{sec:orton} (Eq. \eqref{eq:Gmod}).
    \item By scaling $G(k)$ and picking $C$, obtain normalized basis functions \eqref{eq:singlecurlxi}.
    \item The state is given by \eqref{eq:etastate}.
\end{enumerate}

Rather than \eqref{eq:singlecurl}, we may use any number of curl operators applied in succession, as long as the resulting modes remain normalizable, valid constructions analogous to the above. For instance, one may apply an operator $\nabla \cross \lp( \frac{i}{c}\frac{\partial}{\partial t}\ + \nabla \cross \ \rp)$. In this case $f(r, t) \vec{\hat z}$ mimics a Hertz potential, and the result is a Riemann-Silberstein vector \cite{bialynicki-birula1998}, whose polarization structure is that of a circularly polarized $\vec E$-field. In the remainder of this article the single curl construction is used.

\subsection{Numerical examples}\label{sec:numexample}

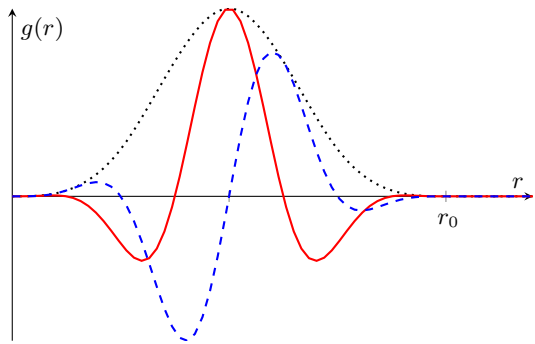
\begin{figure}[tb]
	\begin{tikzpicture}[scale=1]
	\begin{axis}[axis lines=center, xlabel=$r$, ylabel={$g(r)$}, xmax=1.2,xticklabels=\empty,xtick={0.5,1},ytick=\empty,width=8.5cm,height=6cm]
	\addplot[dotted,thick] table [x index=0,y index=1,col sep=comma] {exampleg.csv};
	\addplot[color=red,thick] table [x index=0,y index=2,col sep=comma] {exampleg.csv};
	\addplot[color=blue,thick, dashed] table [x index=0,y index=3,col sep=comma] {exampleg.csv};
	\end{axis}
	\draw (5.8,1.6) node {$r_0$};
	\end{tikzpicture}
	\caption{\label{fig:exampleg} 
		The seed function \eqref{eq:doubletri} which is used in our examples is in the form of an approximately Gaussian envelope with standard deviation $\sigma \approx 0.15 r_0$ (dotted black) multiplied with a harmonic phase factor. Its real and imaginary parts are shown with solid red and dashed blue curves, respectively.}
\end{figure}

\begin{figure}
    \includegraphics{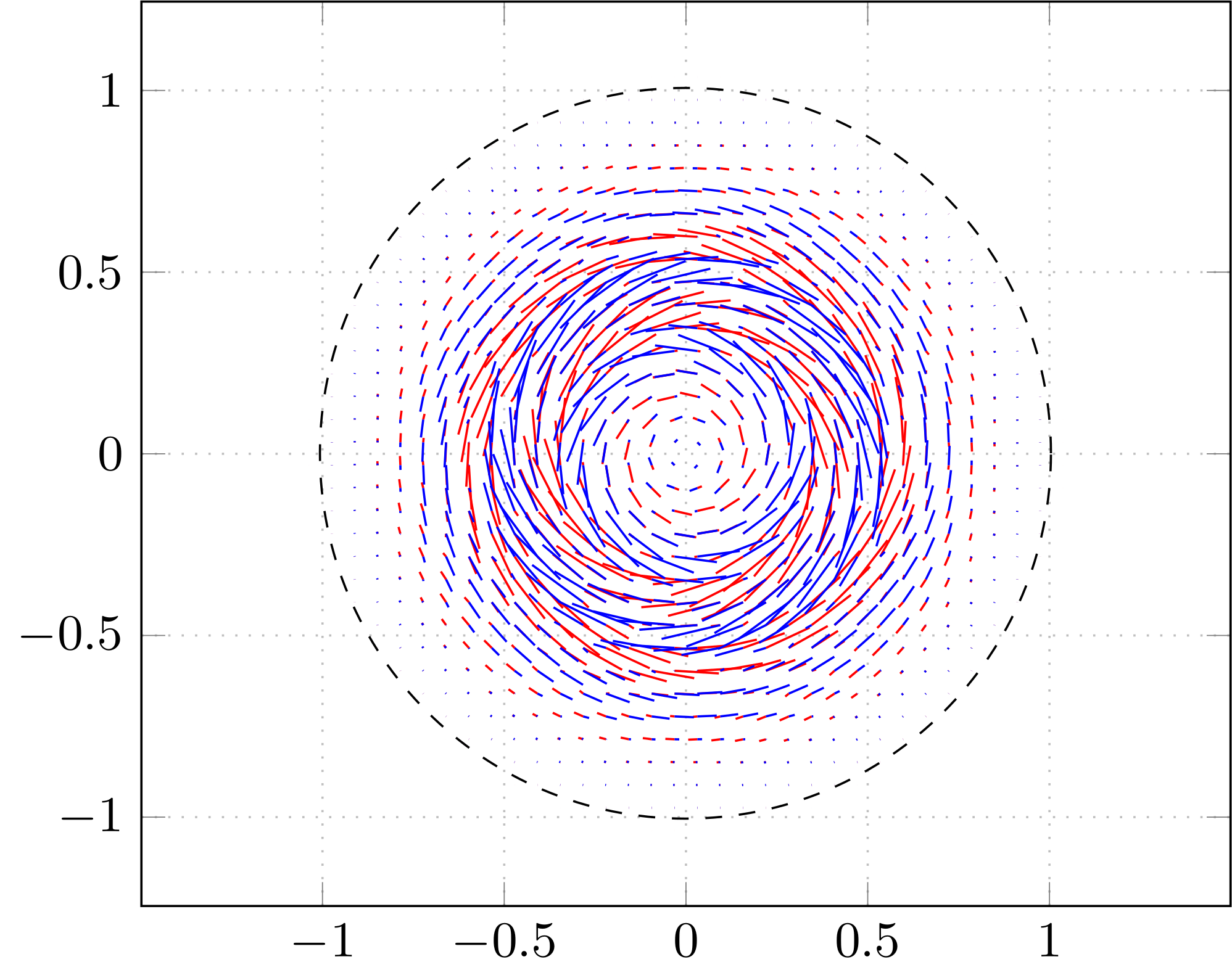}
    \caption{\label{fig:examplef1}
        The transversal generating function $\vec f( \vec r, 0)$ constructed from \eqref{eq:doubletri} according to the single curl construction, shown in the $x$-$y$ plane (arb. units). Red vectors are real components, blue vectors are imaginary components. Dashed line borders the support $V(0)$.}
\end{figure}

\begin{figure}[tb]
	\begin{tikzpicture}[scale=1]
	\definecolor{s1}{RGB}{255, 0, 0}
    \definecolor{s2}{RGB}{225, 40, 25}
    \definecolor{s3}{RGB}{209, 70, 60}
    \definecolor{s4}{RGB}{204, 100, 100}
    \definecolor{s5}{RGB}{200, 140, 150}
    \pgfplotscreateplotcyclelist{set1}{
    s5,every mark/.append style={fill=s1},mark=*\\
    s4,every mark/.append style={fill=s2},mark=*\\
    s3,every mark/.append style={fill=s3},mark=*\\
    s2,every mark/.append style={fill=s4},mark=*\\
    s1,every mark/.append style={fill=s5},mark=*\\
    }
	\begin{axis}[axis lines=center, xlabel=$r$, ylabel={$\exv{\norder{E^2}}/\sin^2 \theta$}, xmax=2, ymax=840, xticklabel style={anchor=north},xticklabels=\empty, xtick={1, 2}, ytick=\empty, cycle list name=set1, no markers]
	\addplot+[thick, fill, fill opacity=0.2] table [x index=0,y index=1,col sep=comma] {expplot_t=0.csv};
	\addplot+[thick, fill, fill opacity=0.2] table [x index=0,y index=1,col sep=comma] {expplot_t=1.csv};
	\addplot+[thick, fill, fill opacity=0.2] table [x index=0,y index=1,col sep=comma] {expplot_t=2.csv};
	\addplot+[thick, fill, fill opacity=0.2] table [x index=0,y index=1,col sep=comma] {expplot_t=3.csv};
	\addplot+[thick, fill, fill opacity=0.2] table [x index=0,y index=1,col sep=comma] {expplot_t=4.csv};
	
	\draw[color=s1] (1.575, 110) node {$t=r_0/c$};
	\draw[color=s5] (0.75, 620) node {$t=0$};
	\end{axis}
	\draw (3.46,-0.25) node {$r_0$};
	\draw (6.9,-0.25) node {$2r_0$};
	\end{tikzpicture}
	\caption{\label{fig:exampleEsquared} 
		The normal ordered squared field expectation value \eqref{eq:E2normalorder} of $\ket{\eta_{1,2}}$ plotted as a function of $r$ for five values of $t$. We have used $\ket{\eta_{1,2}}$ constructed from the seed function \eqref{eq:doubletri} according to the single curl construction. The support $[ct, r_0+ct)$ clearly evolves causally.}
\end{figure}
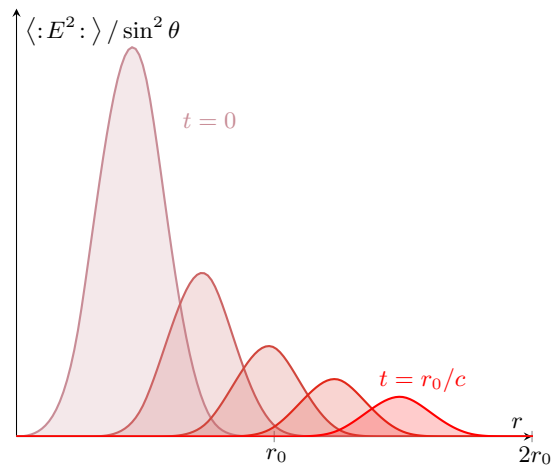

We first consider a state constructed from a seed function 
\begin{equation}
    g(r) = (\text{tri}*\text{tri})\lp(4r/r_0-2\rp)e^{i k_0 r}  \label{eq:doubletri}
\end{equation}
which is supported on $[0, r_0)$ and consists of an envelope multiplied by a complex carrier (see Fig. \ref{fig:exampleg}). Here $\text{tri}(\cdot)$ is the triangular function, which itself is the convolution of a rectangular function $\text{rect}(\cdot)$ with itself. This seed function is chosen because the largest difference between the envelope and its closest Gaussian fit (with standard deviation $\sigma \approx 0.15r_0$) is $1\%$ of the height of the envelope. Thus it is a close approximation to a Gaussian pulse while, as required, being finitely supported.
The vector function $\vec f(\vec r, t)$ corresponding to this seed function (modified according to Appendix \ref{sec:orton}) is shown in \fref{fig:examplef1} at time $t=0$. We have chosen carrier wavenumber $k_0=4\pi/r_0$.
Our single curl construction \eqref{eq:singlecurl} is separable; one may write 
\begin{equation} \label{eq:singlecurlseparate}
    \vec f(\vec r, t) = i\sin \theta \frac{\partial f(r, t)}{\partial r} \hat{\vec{\phi}},
\end{equation}
where $\hat{\vec{\phi}}$ is the unit vector in the $\phi$-direction. In \fref{fig:exampleEsquared} the normal ordered expectation value of the squared field has been plotted as a function of $r$ for five $t$ between $0$ and $r_0/c$. Each curve is supported on $[ct, r_0+ct)$.

It is of interest to quantify how the fidelity \eqref{eq:Fid} changes as a function of localization volume. We note first that coordinate scaling leaves the fidelity invariant. Introducing a coordinate scaled seed function $g(sr)$ where $s$ is a dimensionless scaling factor, a substitution results in a simple scaling of $\xi_1(\vec k)$ and $\xi_2(\vec k)$ which is eliminated by normalization. Clearly then, reducing the pulse width has an identical effect on the fidelity as reducing the carrier wavenumber $k_0$ by the same factor. Because changing the carrier is equivalent to shifting the Fourier transform along the $k$-axis, we quantify how the fidelity changes under shifts of $G(k)$.

The unnormalized Fourier transform of the seed function \eqref{eq:doubletri} is given by
\begin{equation} \label{eq:doubletrifourier}
    G(k) = \text{sinc}^4\lp[(k-k_0)r_0/8\rp]e^{-ikr_0/2},
\end{equation}
which by \eqref{eq:etaG} leads to
\begin{equation}
    \eta(k_0r_0) = \frac{\int_{-\infty}^0 \text{sinc}^8\lp[u-k_0r_0/8\rp] |u| du}{\int_{-\infty}^\infty \text{sinc}^8\lp[u-k_0r_0/8\rp] |u| du}.
\end{equation}
It is clear from this expression that the larger the carrier wavenumber is, the smaller the negative wavenumber integral is, and by extension $\eta$. Thus, as expected, the fidelity tends to $1$ as $k_0r_0 \to \infty$. This will always be the case; introducing a positive carrier wavenumber to $g(r)$ makes the norm in the numerator smaller in proportion to the denominator, improving the fidelity. Naturally, the same applies to increasing the envelope width of $g(r)$.

\begin{figure}[tb]
	\begin{tikzpicture}[scale=1]
	\begin{semilogyaxis}[axis lines=center, xlabel=$k_0^{\text{eff}}r_0/2\pi$, ylabel={$1-F$}, ymax=1, ymin=1e-11, xticklabel style={anchor=north},  y label style={at={(-0.11,1.07)}}, xmin=0, xmax = 8, minor x tick num=1]
	\addplot[color=red,semithick] table [x index=0,y index=1,col sep=comma] {Fidtwiddle.csv};
	\addplot[color=blue,semithick, dotted] table [x index=0,y index=1,col sep=comma] {Fidgaussexact.csv};
	\addplot[color=blue,semithick] table [x index=0,y index=1,col sep=comma] {Fidtwiddlegauss.csv};
	\end{semilogyaxis}
	\end{tikzpicture}
	\caption{\label{fig:fid}
	    $1-F$ as a function of $k_0^{\text{eff}}r_0$ for two $\ket{\eta_{1,2}}$-states constructed from the near-Gaussian \eqref{eq:doubletri} (in red), and the truncated Gaussian \eqref{eq:truncgauss} with $\sigma=r_0/8$ (in blue) using the single curl construction. The quantity $k_0^{\text{eff}}r_0/2\pi$ corresponds to the number of cycles in the pulse, where $k_0^{\text{eff}}$ is the effective carrier wavenumber as discussed in the main text. The fidelity of these states is quite close to $1$ with $\sim 2$ cycles within the localization interval $(0,r_0)$, and improves rapidly with increasing number of cycles.
	    The dotted blue curve is the analytic solution for the Gaussian seed function, assuming negligible truncation. By comparison we see that the effect of the truncation is to flatten out the curve after some number of cycles.
		}
\end{figure}
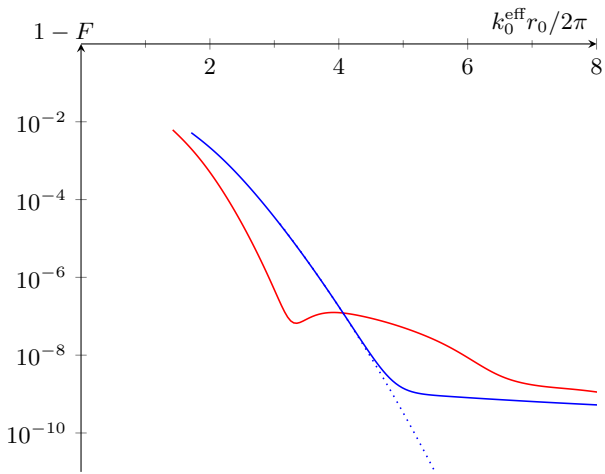


\begin{figure}[tb]
	\begin{tikzpicture}[scale=1]
	\begin{semilogyaxis}[axis lines=center, xlabel=$r_0/\sigma$, ylabel={$1-F_{\text{max}}$}, ymax=1, xticklabel style={anchor=north},  y label style={at={(-0.13,1.075)}}, xmax=4, ymin=0.5e-7]
	\addplot[color=red, semithick] table [x index=0,y index=1,col sep=comma] {Fidupperspherical4withshift.csv};
	\addplot[color=blue, semithick] table [x index=0,y index=1,col sep=comma] {Fidlowersphericalmodified4withshift.csv};
	\end{semilogyaxis}
	\end{tikzpicture}
	\caption{\label{fig:fidupperbound}
	    The narrow-band upper bound \eqref{eq:upperbound} on the fidelity between a given photon and any strictly localized state in a sphere with radius $r_0$ (in red). The photon has a near-Gaussian spectrum \eqref{eq:nearGausstrunc} with carrier $k_0 \sigma=20$ such that its Fourier transform is centered at $r_0/2$ in position space. The fluctuations are caused by lobes formed by interference between inward and outward propagating components. The lower bound \eqref{eq:lowerbound} for the same photon is shown in blue.
		}
\end{figure}
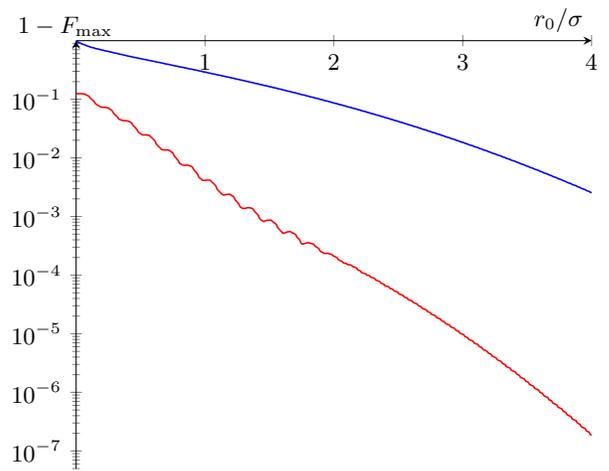

We next consider a truncated Gaussian seed function
\begin{equation} \label{eq:truncgauss}
    g(r) \propto \begin{cases}
    e^{-(r-r_0/2)^2/2\sigma^2} e^{ik_0r}; \ \ \ \ 0 \leq r \leq r_0, \\
    0; \ \ \ \ \ \ \ \ \ \ \ \ \ \ \ \ \ \ \ \ \ \ \ \ \ \ \ \ \text{otherwise.}
    \end{cases} 
\end{equation}
In general, its Fourier spectrum $G(k)$ is found numerically.
In the special case $\sigma\ll r_0$, we can ignore the necessary truncation in $r=0$ and $r=r_0$, which gives $|G(k)|^2 = \exp[-(k-k_0)^2 \sigma^2]$. With the help of \eqref{eq:etaG}, this leads to
\begin{equation}\label{eq:etaanalytical}
    \eta = \frac{1}{2} \lp( 1 - \frac{\sqrt\pi k_0\sigma}{e^{-k_0^2 \sigma^2} + \sqrt\pi k_0\sigma \erf(k_0\sigma)} \rp)
\end{equation}
Eq. \eqref{eq:etaanalytical} allows us to make a comparison to see the effect the truncation has on the fidelity.

After orthogonalizing the above examples (Appendix \ref{sec:orton}), the fidelity always improves further. We identify one reason for this being that the procedure suppresses the negative $k$-spectrum of the seed-function (the central frequency is displaced towards higher values of $k$). This effect has a higher relative impact on the result the closer the location of the peak originally is to $k=0$. Therefore it may be a misleading improvement in the small $k_0$ domain.

In \fref{fig:fid}, we therefore plot $1-F$ as a function of the \emph{effective carrier wavenumber} $k_0^{\text{eff}}$, measured by the average value of $k$ weighted by $|G(k)k|^2$ for positive $k$ (normalized). This corresponds to the average frequency of the modal electric field \eqref{eq:Emode} if the wave is observed at a fixed point sufficiently far away, as can be shown by inserting \eqref{eq:singlecurlxi1} and the associated polarization vector, and take the resulting curl operation outside the integral. To demonstrate the necessity of introducing an effective carrier wavenumber, note that the curves in Fig. \ref{fig:fid} starts at around $k_0^{\text{eff}}r_0/2\pi \sim 2$ (roughly $2$ cycle pulses), even though in terms of the unmodified carrier wavenumber, $k_0r_0\to 0$ here. In other words, for a seed function pulse with less than one cycle, the resulting modal field still has $\sim 2$ cycles within the localization radius. On the other hand, for large values of $k_0^{\text{eff}}r_0$, we have $k_0^{\text{eff}}r_0 \approx k_0r_0$.

\section{Upper and lower fidelity bounds}\label{sec:upper_bound}
We have constructed a class of strictly localized states $\ket{\eta_{1,2}}$ which are close to single photons, as measured by the fidelity \eqref{eq:Fid}. Our goal is now to find upper and lower bounds on the fidelity, comparing any state strictly localized to $V$ with a given single photon, specified with wavenumber spectrum $\xi(\vec k)$ and polarization $\vec e_1(\vec k)$:
\begin{equation}\label{eq:1xi}
    \ket{1_{\xi}} = a_\xi^\dagger \ket{0}, \tbuff a_\xi^\dagger = \int \xi(\vec k) a_1^\dagger(\vec k) d^3k. 
\end{equation}
The maximum fidelity is defined by
\begin{equation}\label{eq:Fmax}
    F_{\max} = \max_{\ket{\psi}}|\braket{1_{\xi}}{\psi}|,
\end{equation}
where $\ket{\psi}$ is a strictly localized state localized to a given volume $V$. 

For the upper bound, we will build our derivation on a bound for one-sided time-localized states \cite{gulla2021b}. The idea is to exploit the fact that the trace distance (and therefore the fidelity) between two quantum states quantify the ability to distinguish between them. To this end, we form an observable which is local to the complement of $V$, which can be used to distinguish between the given single photon state and any strictly localized state. Indeed, since a strictly localized state gives the same result as for vacuum, while the single photon state has a tail outside $V$, the measurement statistics will be different.

The 3D case will, however, turn out to be quite different from the 1D case. Unlike the 1D case, we are not able to obtain an explicit expression for the bound in terms of the photon tail; the bound must be evaluated numerically in the general case. In the special case where the photon has a narrow band of frequencies, we will obtain a closed-form expression for the bound in a similar way as for the exact 1D case. 

For the sake of brevity, the following analysis concerns itself with deriving a suitable, local observable and its properties, which cannot be done similarly to the 1D case. Once we have described the observable and obtained \eqref{eq:axiazeta}, the result from \cite{gulla2021b} applies readily.

The electric field operator $\vec E(\vec r,t)$ is given by the conventional expression \eqref{eq:Efieldop}, where we take $\mathcal E(\omega)\propto\sqrt{\omega}$ to be positive. A proper observable $E_\zeta$ is obtained by smearing,
\begin{equation}
E_\zeta = \frac{1}{4\pi^{3/2}} \int \vec \zeta(\vec r) \cdot \vec E(\vec r, 0) d^3r,
\end{equation}
where $\vec \zeta(\vec r)$ is a real smearing (vector) function. Since we want $E_\zeta$ to indicate a difference between a strictly localized state in $V$ and a single photon, we require $\vec\zeta(\vec r)$ be supported outside of $V$. The smeared field can be expressed
\begin{equation}
	E_\zeta = \frac{1}{\sqrt 2}\lp( a_\zeta + a_\zeta^\dagger \rp),
\end{equation}
where
\begin{equation}\label{eq:azeta}
a_\zeta^\dagger = \sum_{i=1}^2 \int \mathcal{E}(\omega) \vec e_i^*(\vec k) \cdot \vec\zeta(\vec k) a_i^\dagger(\vec k) \dd^3 k,
\end{equation}
and $\vec\zeta(\vec k)$ is the Fourier transform of $\vec\zeta(\vec r)$. We normalize the smearing function such that
\begin{equation}\label{eq:normzeta}
	\sum_{i=1}^2 \int \mathcal E^2(\omega) \lp| \vec e_i^*(\vec k)\cdot \vec\zeta(\vec k) \rp|^2 \dd^3 k = \int \mathcal E^2(\omega) \lp| \vec\zeta(\vec k) \rp|^2 \dd^3 k = 1.
\end{equation}
Then $\comm*{a_\zeta}{a_\zeta^\dagger} = 1$, and the smeared field observable $E_\zeta$ is analogous to the position operator for a regular quantum harmonic oscillator.

An important quantity turns out to be 
\begin{equation}\label{c_xi}
c_\xi = \int \mathcal E(\omega) \vec\zeta^*(\vec k) \cdot \vec e_1(\vec k) \xi(\vec k) \dd^3 k.
\end{equation}
With the help of the Plancherel theorem, \eqref{c_xi} can be rewritten to
\begin{equation}\label{c_xip}
    c_\xi = \int_{\vec r\notin V} \vec \zeta(\vec r) \cdot \vec E_\xi(\vec r)  \dd^3 r,
\end{equation}
with
\begin{equation}\label{eq:Exi}
    \vec E_\xi(\vec r) = \frac{1}{(2\pi)^{3/2}} \int \mathcal E(\omega)\xi(\vec k)\vec e_1(\vec k) e^{i\vec k\cdot\vec r} \dd^3 k.
\end{equation}

To invoke the result in \cite{gulla2021b}, we need to decompose $a_\xi^\dagger$ into a term proportional to $a_\zeta^\dagger$ and a term which commutes with $a_\zeta$. To this end, rewrite
\begin{equation}\label{eq:azetaxi}
    a_\zeta^\dagger = c_1 a_{\zeta1}^\dagger + c_2 a_{\zeta2}^\dagger,
\end{equation}
where $c_1 a_{\zeta1}^\dagger$ and $c_2 a_{\zeta2}^\dagger$ are the two terms in \eqref{eq:azeta}, respectively, and $c_1$ and $c_2$ are normalization constants such that $a_{\zeta1}^\dagger$ and $a_{\zeta2}^\dagger$ are ladder operators. Note that $c_1$ and $c_2$ can be taken real, and $c_1^2 + c_2^2 = 1$. By calculating the component of $\xi(\vec k)$ parallel to $\mathcal{E}(\omega) \vec e_1^*(\vec k) \cdot \vec\zeta(\vec k)/c_1$, we can decompose $a_\xi^\dagger$ as follows:
\begin{equation}\label{eq:axiazeta}
\begin{aligned}
    a_\xi^\dagger &= \int \xi(\vec k) a_1^\dagger(\vec k) d^3k = \frac{c_\xi}{c_1} a_{\zeta1}^\dagger \,+ \perp \\
    &= c_\xi c_1 a_{\zeta1}^\dagger + c_\xi c_2 a_{\zeta2}^\dagger - c_\xi c_1 a_{\zeta1}^\dagger - c_\xi c_2 a_{\zeta2}^\dagger + \frac{c_\xi}{c_1}a_{\zeta1}^\dagger \,+ \perp \\
    &= c_\xi a_\zeta^\dagger \,+ \perp,
\end{aligned}
\end{equation}
where $\perp$ is short-hand for terms that commute with $a_\zeta$. 

Note that $E_\zeta$ is an observable local to the complement of $V$. Moreover, we have the connection between the ladder operators associated with $\zeta$ and $\xi$ in \eqref{eq:axiazeta}. With these ingredients, we can use the result (Eq. (63) from Ref. \cite{gulla2021b}), which states that the fidelity is bounded by
\begin{equation}\label{Fmax_1xi_c_xi}
F_{\text{max}} \leq \sqrt{1 - \frac{2}{\pi e} \abs{c_\xi}^4}.
\end{equation}

It remains to select a suitable smearing function for our observable. We choose to define
\begin{equation}\label{choiczeta}
\vec \zeta(\vec r) =
\begin{cases}
0; & \vec r \in V, \\
f_0 e^{i\phi/2}\vec E_\xi(\vec r) + f_0 e^{-i\phi/2}\vec E_\xi^*(\vec r); & \vec r \notin V,
\end{cases}
\end{equation}
for some normalization constant $f_0>0$ and real phase $\phi$. 
Clearly the smearing function $\vec\zeta(\vec r)$ is supported outside $V$, and real, as required. From \eqref{c_xip} we obtain
\begin{equation}\label{eq:cfzetan}
\begin{aligned}
c_\xi &= f_0 e^{-i\phi/2} \int_{\vec r\notin V} |\vec E_\xi(\vec r)|^2 \dd^3 r \\
&+ f_0 e^{i\phi/2} \int_{\vec r\notin V} \vec E_\xi^2(\vec r) \dd^3 r.
\end{aligned}
\end{equation}
The upper bound is now given by \eqref{Fmax_1xi_c_xi} with \eqref{eq:cfzetan}, subject to normalization \eqref{eq:normzeta}. To make the bound as tight as possible, $\phi$ is picked to maximize $|c_\xi|$. 

The upper bound is quite intuitive; it means that the fidelity between any state strictly localized to $V$ and a single photon is bounded from above by the size of the photon's tail outside $V$.

In the special case where the given photon has a narrow band of frequencies, we can approximate $\mathcal E(\omega)$ as a constant $\mathcal E$ in the normalization integral \eqref{eq:normzeta}, and use the Plancherel theorem to obtain
\begin{equation}\label{eq:normzetak0}
	\mathcal E^2 \int_{\vec r \notin V} \lp| \vec\zeta(\vec r) \rp|^2 \dd^3 r \approx 1.
\end{equation}
Under the same approximation, we have
\begin{equation}
    \vec E_\xi(\vec r) \approx \mathcal E \vec\xi(\vec r),
\end{equation}
with 
\begin{equation} \label{eq:vecxi}
\vec\xi(\vec r) = \frac{1}{(2\pi)^{3/2}} \int \xi(\vec k)\vec e_1(\vec k) e^{i\vec k \cdot \vec r} d^3k.
\end{equation}
Defining
\begin{subequations} \label{eq:munu}
	\begin{align}
		\mu &= \int_{\vec r \notin V} \lp| \vec \xi(\vec r) \rp|^2 \dd^3 r, \label{eq:mu}\\
		\nu &= \int_{\vec r \notin V} \vec \xi^2(\vec r) \dd^3 r, 	
	\end{align}
\end{subequations}
and determining $f_0$ from \eqref{eq:normzetak0}, we can express \eqref{eq:cfzetan} as
\begin{equation}
\abs{c_\xi}^2 = \frac{1}{2}\frac{\mu^2 + \abs{\nu}^2 + 2 \mu \abs{\nu} \cos(\phi + \theta_\nu)}{\mu + \abs{\nu}\cos(\phi + \theta_\nu)}.
\end{equation}
The phase $\theta_\nu = \arg\nu$ is given, but we are free to maximize with respect to $\phi$. Setting $\cos(\phi + \theta_\nu) = 1$ gives
\begin{equation}\label{c_xi_mu}
\abs{c_\xi}^2 = \frac{1}{2} \lp( \mu + \abs{\nu} \rp) \geq \frac{\mu}{2}. 
\end{equation}
Inserting \eqref{c_xi_mu} into \eqref{Fmax_1xi_c_xi}, we finally get the upper bound
\begin{equation}\label{eq:upperbound}
F_{\text{max}} \leq \sqrt{1 - \frac{1}{2\pi e}(\mu+|\nu|)^2} \leq \sqrt{1 - \frac{1}{2\pi e}\mu^2}.
\end{equation}

We have calculated the upper bound for a photon with $\xi(\vec k)$ given by \eqref{eq:singlecurlxi1}, where 
\begin{equation}\label{eq:nearGausstrunc}
    G(k) \propto \begin{cases}
    \exp\lp[-\frac{\sigma^2}{2} (k-k_0)^2 \rp] e^{-i k r_0/2}; \ \ \ \ k\geq 0, \\
    0; \ \ \ \ \ \ \ \ \ \ \ \ \ \ \ \ \ \ \ \ \ \ \ \ \ \ \ \ \ \ \ \text{otherwise.}
    \end{cases}
\end{equation}
is a Gaussian spectrum truncated for negative $k$. To lower the effect of the truncation and ensure that the narrow band approximation may apply, it is given a carrier wavenumber such that $k_0 \sigma=20$. The corresponding spatial mode is centered at $r=r_0/2$ due to the harmonic factor $e^{-ik r_0/2}$. This specific function is chosen because Gaussian pulses are common photonic states considered in the literature, and it is therefore of interest to consider precisely how close a given such state can be to a strictly localized state. In \fref{fig:fidupperbound} the upper bound fidelity for this photon in the narrow-band approximation is plotted as a function of the radius $r_0$ of a spherical localization volume $V$. We have used the tightest bound in \eqref{eq:upperbound}, including both $\mu$ and $|\nu|$. Appendix \ref{sec:calcupperbound} outlines the details of how the upper bound calculations have been performed numerically. In the figure we observe that the fidelity $F_\text{max}$ tends quickly to unity as $r_0/\sigma$ increases. This is expected: Eq. \eqref{eq:nearGausstrunc} describes a Gaussian spectrum with a numerically negligible truncation, corresponding to a Gaussian pulse in the spatial domain, and the maximum fidelity is determined by the photon's tail outside the localization region.

A lower fidelity bound is now quite simple to obtain, since we have already constructed the strictly localized example states in Sec. \ref{sec:etastates}. Given a single photon with $\xi(\vec k)$ specified by \eqref{eq:singlecurlxi1} and \eqref{eq:nearGausstrunc} we inverse Fourier transform $G(k)$ to obtain $g(r)$. To obtain a valid seed function, $g(r)$ must be truncated outside $0\leq r \leq r_0$. The result is a valid seed function for the construction in Sec. \ref{sec:etastates}, so we Fourier transform back and calculate the parameter $\eta$ with \eqref{eq:etaG}. The resulting fidelity is given by
\begin{equation}\label{eq:lowerbound}
    F_{\max} = \max_{\ket{\psi}} \abs{\braket{1_{\xi}}{\psi}} \geq \abs{\braket{1_{\xi}}{\eta_{1,2}}} 
    = \abs{\braket{1_{\xi}}{1_1\ 0_2}} F,
\end{equation}
where we have used \eqref{eq:r_state_expansion}. The inner product $\braket{1_{\xi}}{1_1\ 0_2}$ is the overlap between two single photon states; the given single photon state as specified by $\xi(\vec k)$, and the single photon part of the construction in Sec. \ref{sec:etastates} as resulting from the truncated seed function. For larger localization volumes relative to the photon pulse width, when $F \to 1$, this factor dominates and we find that the lower bound is $\sim |\braket{1_{\xi}}{1_1\ 0_2}|$. The lower bound \eqref{eq:lowerbound} is plotted in \fref{fig:fidupperbound}, and compared to the corresponding upper bound.

\section{Conclusion}
We have constructed a class of three dimensional strictly localized states which can be made arbitrarily close to single photons. These states are strictly localized to finite, causally expanding volumes. We have provided expressions for their fidelity, energy density expectation value, and number expectation value. 

We have also derived an upper bound on the fidelity between a given single photon and any state strictly localized to a volume $V$. The bound is expressed in terms of the size of the photon's tail outside $V$. A corresponding lower bound on the maximum fidelity is given by the inner product between the photon and a suitably constructed example of our strictly localized states.

The class of strictly localized states close to single photons, along with the upper bound, constitute a limit of localization which complements the weak localization limit \cite{bialynicki-birula1998}.


\appendix

\section{Orthonormalization}\label{sec:orton}
For a general seed function, as described in the algorithm in Subsec. \ref{sec:constructvecf}, the resulting modes \eqref{eq:singlecurlxi} are not necessarily orthonormal. We will now show how orthogonalization can be achieved, by a minor modification of $G(k)$ and appropriate normalization. The modification leads actually to an improvement of the resulting fidelity \eqref{eq:Fid}.

Given a seed function $g(r)$ and its Fourier transform $G(k)$, let $\xi_1$ and $\xi_2/C$ (unnormalized) be given by \eqref{eq:singlecurlxi}. We do not care what $C$ equals at this point; only $\xi_2/C$ is of importance. Let $(\cdot, \cdot)$ and $\| \cdot \|$ denote the $\mathcal L^2$ inner product and norm, respectively, of functions on $\reals^3$. We define 
\begin{equation}
    \eta = \frac{\|\xi_2/C\|^2}{\|\xi_1\|^2 + \|\xi_2/C\|^2},
\end{equation}
which is consistent with the earlier definition \eqref{eq:eta}. Furthermore, we introduce the quantity
\begin{align}
    I &= \frac{(\xi_1, \xi_2/C )}{\|\xi_1\|^2 + \|\xi_2/C\|^2},
\end{align}
For the following we assume the inner products involved are all finite. We assume $\eta \leq 1/2$. This can always be ensured by relabeling the modes.

We define a modified $G$-function
\begin{equation}\label{eq:Gmod}
    \widetilde{G}(k) = G(k) - \beta G^*(-k),
\end{equation}
where $\beta$ is some complex constant. This corresponds to a modified seed function
\begin{equation}
    \widetilde{g}(r) = g(r) - \beta g^*(r),
    \label{eq:modifiedseed}
\end{equation}
which preserves the localization properties of $g(r)$. The orthogonality condition for the basis functions associated with the new $\widetilde{G}(k)$ is
\begin{equation}
    \beta^2 I^* - \beta + I = 0,
\end{equation}
from which we can find
\begin{equation}
    \beta = \frac{1}{2I^*}(1-J), \tbuff J = \sqrt{1-4|I|^2}.
\end{equation}
The modification \eqref{eq:modifiedseed} will now lead to orthogonal basis functions
\begin{subequations}
    \begin{align}
        \widetilde{\xi_1}(k) &= \frac{\xi_1(k) - \beta \xi_2(k)/C}{\| \xi_1(k) - \beta \xi_2(k)/C \|}, \\
        \widetilde{\xi_2}(k)/\widetilde{C} &= \frac{\xi_2(k)/C - \beta^* \xi_1(k)}{\| \xi_1(k) - \beta \xi_2(k)/C \|}, \label{eq:modxi2}
    \end{align}
\end{subequations}
here both scaled such that $\|\widetilde{\xi_1}\|=1$. The requirement $\|\widetilde\xi_2\| = 1$ then determines $\widetilde C$ from \eqref{eq:modxi2}. This modification leads to a new $\widetilde\eta$ parameter which satisfies
\begin{equation}
    \widetilde{\eta} - \eta = -\frac{1-J}{2J}\lp(1-2\eta\rp).
\end{equation}
Since $\widetilde\eta\leq\eta$, the modification $G(k)\mapsto \widetilde G(k)$ leads to an improved fidelity \eqref{eq:Fid}.

The exception in which this procedure does not work is if the original $g(r)$ leads to parallel $\xi_1$ and $\xi_2$, corresponding to the case $\eta=1/2$ (this happens when $g(r)$ has constant complex phase). Introducing a harmonic phase factor to the seed function, as described in more detail in Sec. \ref{sec:numexample}, can always be done to avoid this problem.

\section{Solving for expectation values and fidelity} \label{sec:etaproperties}
To find good expressions for expectation values and inner products we employ a few useful quantities. The first is the \emph{squeezed vacuum} \cite{schumaker1985new}, given by
\begin{equation}
    S \ket{0} = \frac{1}{\cosh \gamma} \sum_{i=0}^\infty (-\tanh \gamma)^i \ket{i_1, i_2},
\end{equation}
and secondly the \emph{squeezed ladder operators}
\begin{subequations}
\begin{align}
    S a_1 S^\dagger &= a_1 \cosh \gamma + a_2^\dagger \sinh \gamma, \\
    S a_2 S^\dagger &= a_2 \cosh \gamma + a_1^\dagger \sinh \gamma,
\end{align}
\end{subequations}
which produce the \emph{squeezed field operator},
\begin{equation}
\begin{aligned}
    S \vec E S^\dagger &= \vec E_1 (a_1 \cosh \gamma + a_2^\dagger \sinh \gamma) \\
    +& \vec E_2 (a_2 \cosh \gamma + a_1^\dagger \sinh \gamma) 
     +\sum_{i=3}^\infty \vec E_i a_i + \text{h.c.}
\end{aligned}
\end{equation}
With this in mind we can see that the expectation value of the squared field operator for a state $\ket{\eta_{1,2}^n} = (S^\dagger A_1^\dagger S)^n\ket{0}$ can be written as
\begin{equation}
    \bra{\eta_{1,2}^n}\vec E^2\ket{\eta_{1,2}^n} = (\bra{0}S^\dagger)A_1^n (S \vec E S^\dagger)^2 A_1^{\dagger n}(S\ket{0}),
\end{equation}
and so the expression can be reduced to a combination of finite terms which are series of real numbers. These are generally similar to polylogarithmic series, the simplest of which being the geometric series or related to it by a derivative. Using this kind of factorization we reach the main quantities of interest:
firstly the normal ordered field strength
\begin{equation}
\begin{aligned}
    &\bra{\eta_{1,2}^n}\norder{\vec E^2}\ket{\eta_{1,2}^n} = 2|\vec f_1|^2 (n+\cosh (2 \gamma)) \cosh^2 \gamma \\
    &- 2 \Re{ \vec f_{1}^* \cdot \vec f_{1-}} \cosh^2 \gamma - 4\Re{\vec f_1 \cdot \vec f_2}\frac{M_n(\tanh^2 \gamma)}{\tanh \gamma},
\end{aligned} \label{eq:E2normalorder}
\end{equation}
where we have defined the functions
\begin{subequations}
\begin{align}
    \vec f_1              &= \vec E_1 + \vec E_2^*\tanh \gamma, \\
    \vec f_2              &= \vec E_2 + \vec E_1^*\tanh \gamma, \\
    \vec f_{1-}           &= \vec E_1 - \vec E_2^*\tanh \gamma,
\end{align}
\end{subequations}
and the series $M_n(a) = \sum_{i=0}^\infty \sqrt{i(i+n)}a^i$. Every term in \eqref{eq:E2normalorder} is proportional to the generating function, here denoted $\vec f_1$, and is therefore zero outside of the localization as expected.

It is also of interest to consider the number operator
\begin{equation}
    N = \sum_{j=1}^2 \sum_{i=0}^\infty a_{ij}^\dagger a_{ij}
\end{equation}
This is not a local operator, but measures globally the expected number of photons in all modes. We obtain the expectation value
\begin{equation}
\begin{aligned}
    \bra{\eta_{1,2}^n} N \ket{\eta_{1,2}^n} &= \sum_{j=1}^2 \sum_{i=0}^\infty \bra{\eta_{1,2}^n} a_{ij}^\dagger a_{ij} \ket{\eta_{1,2}^n}\\
    &= (n+2 \sinh^2 \gamma)(\cosh^2 \gamma + \sinh^2 \gamma) \\
    &+ 2 \sinh^2 \gamma - 4 H_n(\tanh^2 \gamma).
\end{aligned} \label{eq:expectationnumber}
\end{equation}
Using this same approach \footnote{After resolving the inner products, one will end up with a triple summation with a Kronecker delta in all three summation indices, in addition to $n$. These can be solved by appropriate change of the summation order and limits.} we obtain an expression for the general fidelity between $\ket{\eta_{1,2}^n}$ and its corresponding number state,
\begin{equation}
\begin{aligned}
    \braket{n_1, 0_2}{\eta_{1,2}^n} &=\frac{1}{(\cosh \gamma)^{n+2}} \sum_{i=0}^{\infty} \sqrt{\binom{n+i}{i}} (\tanh^2 \gamma)^i \\
    &= 1-(1+n/2-\sqrt{n+1})\eta + \order{\eta^2}, \label{eq:nFid}
\end{aligned}
\end{equation}
which is shown for the case $n=1$ in \eqref{eq:Fid}.

One can see that \eqref{eq:expectationnumber} approaches $n$ when $\eta\to 0$ (and therefore $\gamma \to 0$), and likewise \eqref{eq:nFid} tends to 1 for all $n$.

\section{Calculating the upper bound} \label{sec:calcupperbound}
For the special case where the given photon has a narrow band of frequencies, the upper bound \eqref{eq:upperbound} can be computed numerically as an effective one-dimensional problem. Here we will outline the method.

The upper bound for the fidelity is expressed in terms of the photon's tail outside $V$. We consider the case where the photon has spectrum (see \eqref{eq:singlecurlxi1}):
\begin{equation}
    \xi(\vec k) = \sin \theta \frac{G(k)}{\mathcal{E}(\omega)},   
\end{equation}
where $G(k)$ is any spectrum truncated for negative $k$. With polarization vector $\vec e_1(\vec k)$ the corresponding, modal electric field \eqref{eq:Emode} is
\begin{equation}\label{eq:Eximode}
    \vec E_\xi(\vec r, t) = \int \vec e_1(\vec k) \sin \theta G(k) e^{i\vec k \cdot \vec r - i\omega t} d^3k.
\end{equation}
Similarly, 
\begin{equation}\label{eq:xiGdecomp}
    \begin{aligned}
    \vec \xi(\vec r) &= \frac{1}{(2 \pi)^{3/2}}\int \vec e_1(\vec k) \xi(\vec k) e^{i\vec k \cdot \vec r} \dd^3k \\
        &= \frac{1}{(2 \pi)^{3/2}} \int \vec e_1(\vec k) \sin \theta \frac{G(k)}{\mathcal{E}(\omega)} e^{i\vec k \cdot \vec r} \dd^3k.
    \end{aligned}
\end{equation}
We identify the form of this function with \eqref{eq:fGdecomp} from our main construction, which we have established that we may write in the form
\begin{equation}
    \vec \xi(\vec r) = \frac{-i}{(2 \pi)^{3/2}}\nabla \times [\xi(r) \hat{\vec{z}}],
\end{equation}
where
\begin{equation}
    \xi(r) = -\frac{2\pi i}{r}\lp[ u(r)-u(-r) \rp]
\end{equation}
is a scalar d'Alembert solution composed of a function
\begin{equation}
    u(r) = \int_0^\infty \frac{G(k)}{\mathcal{E}(\omega)} e^{ikr}\dd k.
\end{equation}
Note that this is equivalent to following our main construction but with a seed function with a different spectrum.
We can reduce the curl operation by using $\xi(r)$'s spherical symmetry, and obtain
\begin{equation}
    \vec \xi(\vec r) = i \sin \theta \frac{\partial \xi(r)}{\partial r} \hat{\vec \phi}.
\end{equation}
This means that \eqref{eq:munu} for the narrow-band approximation \eqref{eq:upperbound} can be reduced to the form
\begin{subequations} \label{eq:munusinglecurl}
	\begin{align}
		\mu &= \int_{\vec r \notin V} \lp| \frac{\partial \xi(r)}{\partial r} \rp|^2 \sin^2 \theta \dd^3 r, \label{eq:musinglecurl}\\
		\nu =& -\int_{\vec r \notin V} \lp( \frac{\partial \xi(r)}{\partial r} \rp)^2 \sin^2 \theta \dd^3 r. 	
	\end{align}
\end{subequations}
Integrating over the angles is trivial, thereby reducing it to a one-dimensional problem numerically. Normalization can be achieved simultaneously by ensuring that \eqref{eq:musinglecurl}, with integration over all space, is unity.

In the case of the exact upper bound, rather than the narrow-band approximation, one may integrate $\frac{1}{(2\pi)^{3/2}} \frac{\partial f(r)}{\partial r}$ rather than $\frac{\partial \xi(r)}{\partial r}$ in \eqref{eq:munusinglecurl}, where $f(r)$ is the scalar function given by \eqref{eq:fg}; however, the result must be normalized by \eqref{eq:normzeta}, for which there is in general no convenient simplification.

%

\end{document}